\documentclass[bibnotes, prl, twocolumn,superscriptaddress]{revtex4}%
\usepackage{epsfig,dsfont,amssymb,amsmath,amsthm,amsfonts,amsbsy,mathrsfs}
\usepackage{graphicx}
\usepackage{bm}
\begin{document}
\title{Storage and retrieval of photons under their mutual interaction in Rydberg medium}
\author{Liu Yang}
\altaffiliation{Contributed equally to this work}
\affiliation{Department of Physics, University of Arkansas, Fayetteville, AR 72701, USA}
\affiliation{College of Physics, Jilin University, Changchun 130012, China}
\affiliation{Center for Quantum Sciences, Northeast Normal University, Changchun 130117, China }
\author{Bing He}
\altaffiliation{Contributed equally to this work}
\affiliation{Department of Physics, University of Arkansas, Fayetteville, AR 72701, USA}
\author{Jin-Hui Wu}
\affiliation{College of Physics, Jilin University, Changchun 130012, China}
\affiliation{Center for Quantum Sciences, Northeast Normal University, Changchun 130117, China }
\author{ Zhaoyang Zhang}
\affiliation{Department of Physics, University of Arkansas, Fayetteville, AR 72701, USA}
\affiliation{Key Laboratory for Physical Electronics and Devices of the Ministry of Education, Xian Jiaotong University, 
Xian 710049, China}
\author{Min Xiao}
\email{mxiao@uark.edu}
\affiliation{Department of Physics, University of Arkansas, Fayetteville, AR 72701, USA}
\affiliation{National Laboratory of Solid State Microstructures and School of Physics, Nanjing University, Nanjing 210093, 
China}

\begin{abstract}
Stopping and regenerating a pair of single-photon pulses at adjacent locations in coherently prepared Rydberg atomic ensembles are significantly affected by their effective interaction mediated by Rydberg excitations, and the similar processes can differ notably from the one exhibiting the common Rydberg blockade as with the stationary propagation of multi-photon light beams in the same medium. Based on the complete dynamics, we reveal the detailed features in such processes by finding how the profiles of the involved quantum fields evolve in various situations. The findings help to determine the proper regimes for implementing photon-photon gates and photon transistors. In addition, we discuss the non-adiabatic corrections associated with quickly changing control fields, and illustrate a method that restores the photon pulses' original amplitude during their retrieval unless they are heavily damped before storage.
\end{abstract}
\maketitle

Since its first experimental observations \cite{EIT1, EIT0}, the phenomenon of electromagnetically induced transparency (EIT) in ensembles of cold Rydberg atoms has attracted extensive studies. Different from the ordinary EIT \cite{rv-eit}, there exists van der Waals (vdW) or dipole-dipole interaction between atoms, which modifies the dispersive and dissipative properties of light, as well as the photon correlations, in Rydberg medium \cite{EIT-0, EIT-1, EIT-2,EIT-3, EIT-4, EIT-5, EIT-6, EIT-7, EIT-8, EIT-a, EIT-9, EIT-10}. These interactions also indicate the possible formation of bound states \cite{bd} and lattice structures \cite{lt} of photons. 
More recent experiments \cite{Ex1, Ex2, Ex3, Ex4, Ex5} have demonstrated numerous interesting features related
to Rydberg-atom-mediated interactions.

The above mentioned long-range interactions make Rydberg EIT medium a promising candidate for quantum information processing with photons. A fascinating application of this kind is implementing a deterministic two-qubit gate, the building block of photonic quantum computing. Apart from the scenarios working with slowly traveling photons \cite{g1,g2,g3, g4,g5}, a gate operation could be performed with stopped photons that are converted to Rydberg atomic excitations, so that the nonlocal interaction between the long-living Rydberg spinwaves achieves much more significant nonlinear phases than those generated by the collisional interaction between atoms \cite{c1, c2}. However, because the vdW interaction also brings about unwanted effects, the previous schemes \cite{p1,p2} in this category require more sophisticated technical steps, while it was not clear whether the simplest way to store and regenerate two photons (without extra procedures) properly under vdW potential could realize a good gate operation. On the other hand, other devices such as switches 
\cite{sw} and transistors \cite{tr1,tr2} should perform a different function that scatters more photons with a stored one. Stopping light in Rydberg medium and/or re-converting Rydberg excitations into light have been demonstrated experimentally \cite{sw, tr1, tr2, d1, d2}, and the vdW interaction potential in the used multi-photon beams is generally high and stable, leading to the well-known blockade effect. Yet the details of the corresponding processes for individual photon pulses under their varying interaction, which are important to the operations of the devices such as photon-photon gates and photon transistors, had not been fully understood. 
The previously unknown features in these processes will be illustrated in this work.

\begin{figure}[b!]
\vspace{-0.2cm}
\centering
\epsfig{file=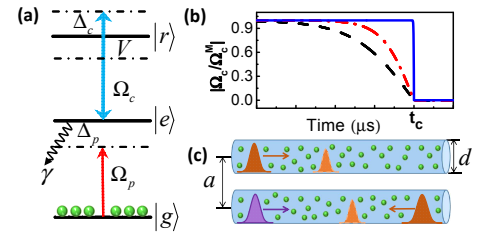,width=0.7\linewidth,clip=}
{\vspace{-0.2cm}
\caption{(a) Atomic level scheme. Here $\Delta_p=\omega_{eg}-\omega_p$, $\Delta_c=\omega_{re}-\omega_c$
(difference between the level gaps and the field frequencies). (b) Control field Rabi frequency $\Omega_c(t)= \Omega^M_{c} \tanh
(t_{c}-t)/\tau_{c} $. Three values, $\tau_{c,1}$ (solid), $\tau_{c,2}$ (dash-dotted) and $\tau_{c,3}$ (dahed), with $\tau_{c,1}<\tau_{c,2}<\tau_{c,3}$, are given as the examples. (c) Geometry of the pulse propagations.  }}
\vspace{-0cm}
\end{figure}

In addition to the application concerns, the main purpose of our study is to provide a more realistic picture for the non-steady motions
of interacting wave packets in Rydberg medium. To reflect the spatial distributions of photons and Rydberg excitations, we treat them 
as delocalized fields for their interactions and find their continuous evolutions in space and time. This differs from most other related studies, which rely on a discontinuous blockade radius from point-point interaction potential.

A typical example of our concerned problem is the setup of two pencil-shaped ensembles in Fig. 1.
Here two single-photon pulses either travel along the same direction (co-propagation) or
respectively enter the opposite tips of the ensembles (counter-propagation). The control fields with the Rabi frequency $\Omega_c(t)$ are being turned off with time, so that the pulses will be stopped and converted to Rydberg excitations that are close to each other. After the two Rydberg spinwave packets interact for a certain time, they will be turned on again to retrieve the photon pulses in the phase matched directions. This process can be modeled by the dynamical evolutions of the electromagnetic fields $\hat{\cal E}_l({\bf x},t)$ of the two photons ($l=1$ and $2$) with the kinetic Hamiltonian $H_p/\hbar=-ic\sum_l\int d{\bf x}\hat{\cal E}^\dagger_l({\bf x})\partial_z\hat{\cal E}_l({\bf x})$, the induced polarization fields $\hat{P}_l({\bf x},t)=\sqrt{N}\hat{\sigma}_{ge}^l({\bf x},t)$ and Rydberg spinwave fields $\hat{S}_l({\bf x},t)=\sqrt{N}\hat{\sigma}^l_{gr}({\bf x},t)$, where
$\hat{\sigma}_{\mu\nu}=|\mu\rangle \langle \nu|$. A high atomic ensemble density $N$ enhances the photon-atom coupling (with the constant $g$) seen in the Hamiltonian $H_{af}/\hbar=-\sum_l\int d{\bf x} \{g\sqrt{N}\hat{\cal E}^\dagger_l({\bf x})\hat{P}_l({\bf x})+\Omega_c(t)\hat{S}^\dagger_l({\bf x})\hat{P}_l({\bf x})+H.c.\}+\sum_l\int d{\bf x}\Delta_p\hat{P}_l^\dagger\hat{P}_l(\bf x) $, 
which describes the atomic level scheme. A narrow band photon propagates with negligible absorption under the EIT condition $\Delta_p+\Delta_c=0$ [$\Delta_{p/c}$ is defined in Fig. 1(a)]. However, under the spinwave interaction with the Hamiltonian $H_{int}=\int d{\bf x}\int d{\bf x'}\hat{S}^\dagger_1({\bf x})\hat{S}_2^\dagger({\bf x'})\Delta({\bf x}-{\bf x'})\hat{S}_2({\bf x'})\hat{S}_1({\bf x})$, where $\Delta({\bf x}-{\bf x'})=C_6/|{\bf x}-{\bf x'}|^6$ is the vdW potential, the condition will be violated by a shift $V$ of the levels $|r\rangle$ in the relevant atoms. The consequent dissipation from populating the levels $|e\rangle$ that decay at the rate $\gamma$ can be depicted by a stochastic Hamiltonian $H_{dis}/\hbar=i\sqrt{2\gamma}\sum_l\int d{\bf x} \{\hat{\zeta}^\dagger_l({\bf x},t)\hat{P}_l({\bf x})-H.c.\}$ involving the quantum noise fields $\hat{\zeta}_l({\bf x},t)$ of 
the reservoirs \cite{g5}.

\begin{figure}[b!]
\vspace{-0cm}
\centering
\epsfig{file=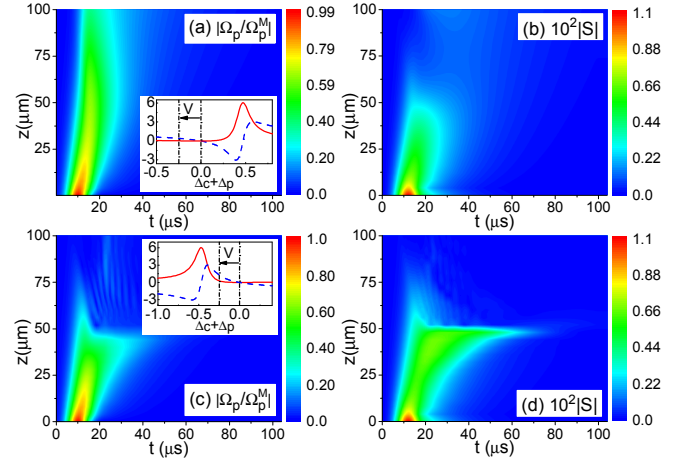,width=1.0\linewidth,clip=}
{\vspace{-0.6cm}
\caption{Dynamics of the photon pulses (a, c) and the induced spinwave in the unit $\mu$m$^{-3/2}$ (b, d), which counter-propagate in two $100$-$\mu$m long ensembles of $^{87}$Rb atoms subject to a control field of $\tau_{c}=1$ $\mu$s, $t_{c}=100$ $\mu$s, and $\Omega_{c}^{M}=1.5\times2\pi$ MHz in Fig.1(b). The field profiles are on the ensemble axis, and obtained from the numerics with the iteration step size $0.002$ $\mu$s in the time direction and $0.02$ $\mu$m in the spatial direction.
Here, $\left\vert g\right\rangle =5S_{1/2}$, $\left\vert e\right\rangle =5P_{1/2}$, and $\left\vert r\right\rangle =100S_{1/2}$, with $ C_{6} \approx -2.3\times 10^{5}$ GHz $\mu$m$^{6}$ and $\gamma=5.75\times2\pi$ MHz. We set $\Delta_{p}=-\Delta
_{c}=5\gamma$ in (a, b) and $\Delta_{p}=-\Delta_{c}=-5\gamma$ in (c, d). The photons are Gaussian pulses with $\Omega_{p}^{M}=0.01$ MHz, $t_{p}=10.0$ $\mu$s and $\tau_{p}=5.0$ $\mu$s. The system parameters are $N=2\times10^{13}$ cm$^{-3}$, $a=6$ $\mu$m and $d=2$ $\mu$m. The insets are the corresponding imaginary (solid) and real (dashed) parts of the normalized susceptibility to the integers. }}
\vspace{-0cm}
\end{figure}

It is sufficient to find the mean values $O_l({\bf x},t)=\langle \hat{O}_l({\bf x},t)\rangle$ of the quantum fields $\hat{O}_l({\bf x},t)$, and their evolutions under the total Hamiltonian $H=H_p+H_{af}+H_{int}+H_{dis}$ are determined by the following equations 
(see Appendix A and we set $\hbar=1$ here):
\begin{equation}
\partial_t{\cal E}_l({\bf x},t)+c\partial_z {\cal E}_l({\bf x},t)=ig\sqrt{N}P_l({\bf x},t);
\vspace{-0.2cm}
\label{1}
\end{equation}
\begin{eqnarray}
\partial_t P_l({\bf x},t)&=&-(\gamma+i\Delta_p)P_l({\bf x},t)+i\Omega^\ast_c(t)S_l({\bf x},t)\nonumber\\
&+&ig\sqrt{N}{\cal E}_l({\bf x},t);
\label{2}
\end{eqnarray}
\begin{equation}
\partial_t S_l({\bf x},t)=-iV_l({\bf x},t)S_l({\bf x},t)+i\Omega_c(t)P_l({\bf x},t),
\label{3}
\vspace{-0.3cm}
\end{equation}
where
\begin{equation}
\vspace{-0cm}
V_l({\bf x},t)=\int d{\bf x'} \frac{C_6} {|{\bf x}-{\bf x'}|^6} |S_{3-l}({\bf x'},t)|^2
\label{v}
\end{equation}
for $l=1$ and $2$. Using Eqs. (\ref{2}) and (\ref{3}), one can expand the right-hand side of Eq. (\ref{1}) as
\begin{eqnarray}
&& ig\sqrt{N}P_l({\bf x},t)
=-\frac{g^2 N}{|\Omega_c(t)|^2}\frac{\partial}{\partial t}{\cal E}_l({\bf x},t)\nonumber\\
&-&\frac{g^2 N{\cal E}_l({\bf x},t)}{\Omega_c(t)}\frac{\partial}{\partial t}\frac{1}{\Omega_c^\ast (t)}
+\frac{gN}{\Omega_c(t)}\frac{\partial}{\partial t}\big\{\frac{1}{\Omega^\ast_c(t)}\big(\frac{\partial}{\partial t}+\gamma
\nonumber\\
&+&i\Delta_p\big) \frac{1}{\Omega_c(t)}\big(\frac{\partial}{\partial t}+iV_l({\bf x},t)\big)\frac{g{\cal E}_l({\bf x},t)}{\Omega^\ast_c(t)}\big\}+\cdots
\label{ex}
\end{eqnarray}
Then, in our concerned regime realizing slow light ($g^2 N/|\Omega_c|^2\gg 1$), the time derivative on the left-hand side of Eq. (\ref{1}) will be absorbed into the leading term in the above equation, reducing Eq. (1) to the one only with a spatial derivative. The discretization of this rearranged form of Eq. (1) is used to get the spatial profile of ${\cal E}_l$ at a specified moment, given the values of $P_l$ induced by the field ${\cal E}_l$ at the preceding moment of iteration with the discretized Eqs.(\ref{2}) and (\ref{3}). We apply a second order Runge-Kutta method in the iterative steps toward the quantities (${\cal E}_l$, $P_l$, $S_l$ and $V_l$) all over the space-time grids. For simplicity, only symmetric propagations in two ensembles will be discussed below, but the generalization to different pulse motions is straightforward.

\begin{figure*}[t!]
\vspace{-0cm}
\centering
\epsfig{file=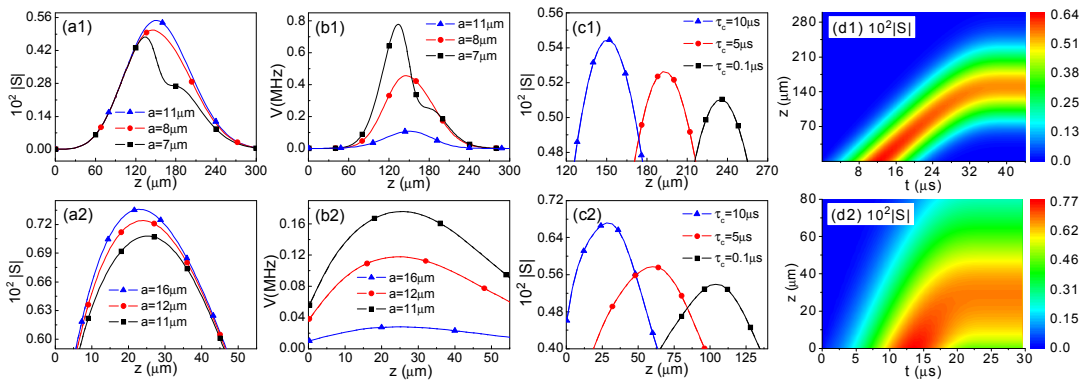,width=0.8\linewidth,clip=}
{\vspace{-0.3cm}
\caption{(a1) Realized spinwave profiles (on the ensemble axis) from stopping two counter-propagating pulses in two $300$-$\mu$m long ensembles, given the different ensemble separations and a switching of $\tau_c=10$ $\mu$s, $t_c=45$ $\mu$s. (b1) Corresponding potentials $V(z)$ the standstill spinwave packets in (a1) create on the other ensemble. (a2) and (b2) show the similar quantities for the co-propagating setups, while the control field is turned off at $t_c=30$ $\mu$s. Ensemble length for co-propagation is flexible as long as the together pulses can be contained inside. A nearly homogeneous potential around $0.02$ MHz is realized with the largest $a$ in (b2). (c1) Spinwave profiles from counter-propagating pulses stopped at different switching speeds 
($a=10$ $\mu$m); (c2) similar profiles for a co-propagating setup. (d1) Example of counter-propagating spinwave dynamics, given $a=10$ $\mu$m, $\tau_{c}=10$ $\mu$s, and $t_{c}=40$ $\mu$s. (d2) Example of co-propagating spinwave dynamics, which differs from (d1) by $t_{c}=24$ $\mu$s. These examples demonstrate that different propagation geometry can achieve similar results, but the co-propagagation setups enjoy the advantage of much shorter ensembles. The common parameters are $\Omega_{c}^{M}=2\times 2\pi$ MHz, $\Omega_{p}^{M}=0.01$ MHz, $\tau_{p}=7$ $\mu$s, $t_{p}=12$ $\mu$s, $N=2\times10^{13}$ cm$^{-3}$, and $d=2$ $\mu$m. The iteration step sizes in the numerical simulations are the same as those in Fig. 2.}}
\vspace{0.2cm}
\end{figure*}

As a time-dependent boundary condition for pulse evolution, the electromagnetic field of the input photon pulses is supposed to have the profile $\Omega_p(\rho, t)=\Omega_p^M e^{-(t-t_p)^2/\tau_p^2}J_0(2\nu_{01}\rho/d)$ at the entries of the ensembles, where $\Omega_p^M$ is the maximum of the photons' Rabi frequency $\Omega_p=g{\cal E}$, and $t_p$ and $\tau_p$ are the time scales indicating the peak arrival and pulse duration, respectively, while $J_0(x)$ is the Bessel function of order zero with its first zero point $\nu_{01}$. The effects of a gradually increasing interaction can be best seen from stopping two counter-propagating pulses. In reality, storage of photons may fail as shown in Fig. 2, which illustrates the associated dynamical evolutions for the photons with the opposite-sign detuning values. There a positively detuned photon will go through the medium, though it is damped. If the photon's detuning changes the sign, it will not survive the travel. Such difference can be explained with the susceptibility $\chi(\omega_p)$, defined as $P(\omega_p)=\sqrt{N}\chi(\omega_p)\Omega_p(\omega_p)$, for a single-frequency wave under a constant interaction potential $V$; see the insets of Fig. 2. In the former situation the attractive potential $V$ ``pulls" the central frequency component initially under the EIT condition away from the absorption peak, but it ``pushes" the corresponding component in the latter toward the peak. Then the positively detuned photons will virtually see two-level systems when they get closer, but the same medium becomes opaque to the negatively detuned ones. These effects are more significant for a higher detuning $\Delta_p$, as the gap between the absorption peak and the two-level regime narrows down accordingly. The scenario of negative detunings is most suitable for implementing a photon transistor. In the presence of many photons at the same time, a significant interaction potential can immediately ``push" the system across the absorption peak to the two-level regime on the other side, fitting the Rydberg blockade model (super-atom) irrespective of the photons' detuning signs \cite{EIT-2} which well describes the stationary propagation of light in Rydberg medium. By contrast with the phenomenological models, our first-principle calculations provide the understandings of what happen to individual photons under their constantly varying interaction and a time-dependent $\Omega_c(t)$ simultaneously.

In order to realize their storage, the photons, as well as the control field, should be better to be resonantly coupled to the energy levels ($\Delta_p=\Delta_c=0$) so that the farther positions of absorption peaks leave more room for the frequency components not to enter the above mentioned regimes. Certainly one can take a stronger control field to widen the EIT window, but the practice suppresses the induced spinwave, as seen from the relation $S\sim g\sqrt{N}{\cal E}/\Omega_c$ for a single-frequency wave without interaction. Accordingly the photon pulses should have a sufficiently long duration $\tau_p$ (corresponding to a narrower bandwidth $\delta \omega_p=1/\tau_p$) to fit into the limited EIT window. Such spatially extending photon pulses will be mapped to Rydberg spinwaves distributing over the ensembles, after the control field is turned off. When the process happens to two pulses counter-propagating respectively in close ensembles, the mutual interaction can diminish their fronts before they are stopped side by side (the interaction between their back portions is still insignificant at this time), leaving a spinwave packet of asymmetric shape as in Fig. 3(a1). The potential $V(z)$ acting on other ensemble can therefore be extremely inhomogeneous as shown in Fig. 3(b1). For an ideal photonic gate operation, each point on the stored wave packets should be under the same potential value $V$, so that the wavefunction will gain a uniform phase. Thus one has to face a trade-off between the magnitude and the uniformity of the achieved potential $V(z)$.

Unlike most other previous researches, we adopt the potential field in Eq. (\ref{v}), which is generated by the real-time spinwave field $S({\bf x},t)$, for the the mutual interaction between Rydberg excitations. The evolution an input pulse undergoes under this potential is very different from that under the interaction with a single-point Rydberg atom. The interaction potential value $V({\bf x}, t)$ is simply the degree of deviating the EIT condition $\Delta_p+\Delta_c=0$ at any space-time point $({\bf x}, t)$, when the control field is on. The consequent space-time dependent dissipation can be captured by our simulations directly based on Eqs.(\ref{1})-(\ref{3}). More examples of such pulse dissipation are given in Appendix B. The calculations also specify the proper regime of larger ensemble separations, in which an approximately homogeneous potential can be realized for gate operation; see Fig. 3(b2). In this regime beyond certain ensemble separation, co-propagation and counter-propagation actually make little difference.

\begin{figure}[b!]
\vspace{-0cm}
\centering
\epsfig{file=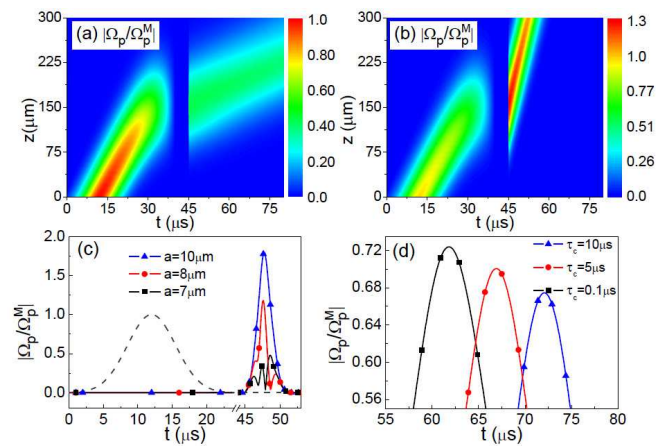,width=1.0\linewidth,clip=}
{\vspace{-0.3cm}
\caption{Dynamical evolutions of photon pulses throughout the whole ``write-in" and ``read-out" process. The storage period is the same as in Fig. 3(d1), with $\Omega_c^{in, M}=2\times2\pi$ MHz, $\tau_c=10$ $\mu$s and $a=10$ $\mu$m. The control fields for retrieval have their Rabi frequencies $\Omega^{out}_c(t)$ as the time-reversals of the functions in Fig. 1(b). We take $\Omega_{c}^{out,M}=1\times2\pi$ MHz in (a) and $\Omega_{c}^{out, M}=3\times2\pi$ MHz in (b), while the switch-on occurs at $\tau_{c}=0.1$ $\mu$s in both cases. (c) Profiles of the retrieved photons for different ensemble separations, where $\Omega_{c}^{out, M}=5\times2\pi$ MHz and $\tau_{c}=0.1$ $\mu$s. The dashed curve is that of the input photons for comparison. (d) Profiles of the photons regenerated with different switching speeds as shown at the exits, given $\Omega_{c}^{out, M}=2\times 2\pi$ MHz and $a=10$ $\mu$m.  }}
\vspace{-0cm}
\end{figure}

The pulse motion in ensemble is largely influenced by the control field. Given the Rabi frequencies in Fig. 1(b), the parameters $t_c$ and $\tau_c$ determine where a pulse will be stopped. A fast switching with small $\tau_c$ can immediately stop the pulses at any moment. One question is whether the non-adiabatic corrections \cite{f-l}, connected with terms containing the time derivatives of $\Omega_c(t)$ on the right-hand side of Eq. (\ref{ex}), could lead to considerable pulse losses. This can be clarified by our numerical calculations with arbitrary $\Omega_c(t)$, which go beyond the previously studied adiabatic passages \cite{ad0, ad1, ad2}. The results in Figs. 3(c1) and 3(c2) indicate unremarkable difference between the disparate switching speeds. Further calculations manifest that, in the ensembles of high densities, the non-adiabatic effects are overshadowed by those from the pulse interaction; see Appendix C.

Under any circumstance the photon absorption is inevitable due to pulse interaction, limited EIT window, non-adiabatic effects and others. We here demonstrate a mechanism that compensates for the previous dissipations through photon regeneration. In Fig. 4(a) the control field in the ``read-out" is weaker than the control field for the ``write-in". The amplitude of the regenerated photons is less than that of the input photons, and they exit the medium at a lower group velocity (flatter contour) as well. The retrieval control field in Fig. 4(b) is, on the other hand, stronger than the one used for stopping the photons, so that the re-converted photon pulses become brighter and leave the ensembles at a higher group velocity (steeper contour). Near the entries and exits of this counter-propagation setup (almost no interaction exists there), the pulses can be regarded as dark-state polariton field $\Psi({\bf x}, t)=\cos\vartheta (t){\cal E}({\bf x}, t)-\sin\vartheta (t) S({\bf x}, t)$, where $\vartheta(t)=\arctan g\sqrt{N}/\Omega_c(t)$. 
The electromagnetic field ${\cal E}({\bf x}, t)$ of the pulses therefore takes the form $\cos\vartheta (t_{in})\Psi({\bf x}, t_{in})$ at the beginning and $\cos\vartheta (t_{out})\Psi({\bf x}, t_{out})$ in the end. To maintain their amplitudes, the dissipation that reduces the field $\Psi({\bf x}, t_{in})$ to $\Psi({\bf x}, t_{out})$ can be partly offset by the factor $\cos\vartheta (t)$, though the retrieved pulses are deformed (such deformation does not affect carrying quantum information). The best enhancement rendering $\cos\vartheta (t_{out})= 1$ for the output photons demands an infinite $\Omega_c^{out}$, implying the impossibility of restoring the amplitudes of the severely damped pulses with a realistic $\Omega_c^{out}(t)$ [See Fig. 4(c)]. Because pulses' dissipation is primarily decided by the duration and intensity of their interaction, a quick retrieval outdoes a slow one, contrary to what happens to 
the storage period; compare Fig. 4(d) with Fig. 3(c1). A slow ``write-in" and a fast ``read-out" are thus favored to our concerned processes.

In summary, we have presented the first study of the details in stopping and regenerating individual photon pulses under their mutual interaction in Rydberg EIT medium. Much more complex and varied than those of the stationary propagation of multi-photon beams, several important features in the processes originate from the ever-changing interactions which lead to space-time dependent dissipation. Our simulations provide more realistic pictures of the continuous photon pulse evolutions in space and time, dispensing with a boundary of discontinuity as the blockade radius in phenomenological models. Apart from clarifying the regimes to implement photon transistors and gates, and ascertaining the insignificant role of the non-adiabatic corrections in high density ensembles, we show that the amplitudes of the weakened photon pulses can be enhanced by a stronger retrieval control field. These understandings could be valuable guide for the relevant experimental researches.

L. Y. is supported by the China Scholarship Council. M.X. acknowledges funding support in part from NBRPC
(Grant No. 2012CB921804). J. H. W. is sponsored by NSFC (Grants No. 11174110) and NBRPC (Grants No. 2011CB921603).

\clearpage
\begin{widetext}

\section*{\large Appendices for ``Storage and retrieval of photons under their mutual interaction in Rydberg medium" }

\subsection*{A.~~~~~  Dynamical equations of quantum fields}

\renewcommand{\theequation}{A-\arabic{equation}}
\setcounter{equation}{0}
\renewcommand{\thefigure}{S-\arabic{figure}}
\setcounter{figure}{0}

In our concerned problem, the photons with the electromagnetic fields $\hat{\cal E}_l({\bf x},t)$ are co-propagating or counter-propagating in two ensembles, having their kinetic Hamiltonian (here we give the form of co-propagation)
\begin{eqnarray}
H_{p}/\hbar &=& -ic\sum_{l=1}^2\int d{\bf x}\hat{\cal E}^\dagger_l({\bf x})\partial_z\hat{\cal E}_l({\bf x}).
\label{kinetic}
\end{eqnarray}
Together with the control field of the Rabi frequency $\Omega_c(t)$, they will induce the polarization fields $\hat{P}_l({\bf x},t)$
and the Rydberg spinwave fields $\hat{S}_l({\bf x},t)$ in the medium, as described by the many-body version of the atomic-level-scheme Hamiltonian
\begin{eqnarray}
H_{af}/\hbar &=& -\sum_{l=1}^2\int d{\bf x} \{g\sqrt{N}\hat{\cal E}^\dagger_l({\bf x})\hat{P}_l({\bf x})+\Omega_c(t)\hat{S}^\dagger_l({\bf x})\hat{P}_l({\bf x})+H.c.\}\
+\sum_{l=1}^2\int d{\bf x}\Delta_p\hat{P}_l^\dagger({\bf x})\hat{P}_l({\bf x}),
\label{couple}
\end{eqnarray}
where $g$ is atom-field coupling constant, $N$ the atomic density of the ensembles, and $\Delta_p$ the detuning of the photons.
The Hamiltonian for the interaction between the Rydberg spinwave fields takes the form
\begin{eqnarray}
H_{int}=\int d{\bf x}\int d{\bf x'}\hat{S}^\dagger_1({\bf x})\hat{S}_2^\dagger({\bf x'})\Delta({\bf x}-{\bf x'})\hat{S}_2({\bf x'})\hat{S}_1({\bf x}).
\label{interaction}
\end{eqnarray}
Meanwhile, as the result of the interaction, the population of the intermediate levels spontaneously decaying at the rate $\gamma$ gives rise to the photon dissipation, and we describe it by the stochastic Hamiltonian
\begin{eqnarray}
H_{dis}/\hbar=i\sqrt{2\gamma}\sum_{l=1}^2\int d{\bf x} \{\hat{\zeta}^\dagger_l({\bf x},t)\hat{P}_l({\bf x})-H.c.\}
\label{diss}
\end{eqnarray}
in terms of the coupling between the polarization fields and quantum noise fields, where the random-variable noise operators satisfy $[\hat{\xi}_l({\bf x},t),\hat{\xi}^{\dagger}_m({\bf x}^{\prime},t^{\prime})]=\delta_{lm}\delta({\bf x}-{\bf x}^{\prime})\delta(t-t^{\prime})$. An advantage of this stochastic Hamiltonian approach is that the physical process can be modeled by
the formal unitary evolution operator $U(t,0)={\cal T}\exp\{-i\int_0^t  H(\tau) d\tau\}$ with $H(t)=H_{af}+H_{int}+H_{p}+H_{dis}$, which naturally leads to the Heisenberg-Langevin equations of the relevant quantum fields as shown below.

In deriving the dynamical equations, we write $d\hat{B}_l({\bf x},t)=\hat{\zeta}_l({\bf x},t)dt$, which satisfy the Ito's rules (a generalization from those in \cite{S1}):
\begin{eqnarray}
&& d\hat{B}_l({\bf x},t)d\hat{B}_l({\bf x},t)=0,~~~d\hat{B}^\dagger_l({\bf x},t)d\hat{B}_l^\dagger({\bf x},t)=0,\nonumber\\
&& d\hat{B}_l^\dagger({\bf x},t)d\hat{B}_l({\bf x},t)=0,~~~d\hat{B}_l({\bf x},t)d\hat{B}_l^\dagger({\bf x},t)=dt.
\end{eqnarray}
Then an increment of the polarization fields, for example, will be found as
\begin{eqnarray}
&& d \hat{P}_l ({\bf x},t)=U^\dagger (t+dt,t)\hat{P}_l ({\bf x},t) U(t+dt,t)-\hat{P}_l ({\bf x},t)\nonumber\\
&=&-\frac{i}{\hbar}\big[\hat{P}_l ({\bf x},t), \big(H_{af}+H_{int}+H_{p}\big)dt-i\hbar \sqrt{2\gamma} \int d{\bf x}' \big(d\hat{B}({\bf x}',t)\hat{P}^\dagger_l ({\bf x}',t)-d\hat{B}^\dagger({\bf x}',t)\hat{P}_l ({\bf x}',t)\big)\big ]\nonumber\\
&+& \gamma \int d{\bf x}'\big(2\hat{P}^\dagger ({\bf x}',t)\hat{P}({\bf x},t)\hat{P}({\bf x}',t)-\hat{P}({\bf x},t)\hat{P}^\dagger({\bf x}',t)\hat{P}({\bf x}',t)-\hat{P}^\dagger({\bf x}',t)\hat{P}({\bf x}',t)\hat{P}({\bf x},t)\big),
\end{eqnarray}
where the above Ito's rules have been applied to the expansion up to the second order of $-iH_{dis}(t)dt$. In this way the
exact dynamical equations for the three types of quantum fields are obtained as follows ($\hbar=1$):
\begin{equation}
\partial_t\hat{{\cal E}}_l({\bf x},t)+c\partial_z \hat{{\cal E}}_l({\bf x},t)=ig\sqrt{N}\hat{P}_l({\bf x},t);
\label{a1}
\end{equation}
\begin{eqnarray}
\partial_t \hat{P}_l({\bf x},t)&=&-(\gamma+i\Delta_p)\hat{P}_l({\bf x},t)+i\Omega^\ast_c(t)\hat{S}_l({\bf x},t)
+ig\sqrt{N}\hat{{\cal E}}_l({\bf x},t)-\sqrt{2\gamma}\hat{\zeta}_l({\bf x},t);
\label{a2}
\end{eqnarray}
\begin{equation}
\partial_t \hat{S}_l({\bf x},t)=-i\hat{V}_l({\bf x},t)\hat{S}_l({\bf x},t)+i\Omega_c(t)\hat{P}_l({\bf x},t),
\label{a3}
\end{equation}
where
\begin{equation}
\vspace{-0cm}
\hat{V}_l({\bf x},t)=\int d{\bf x'} \Delta({\bf x}-{\bf x}')\hat{S}^\dagger_{3-l}({\bf x'},t)\hat{S}_{3-l}({\bf x'},t)
\label{inter}
\end{equation}
for $l=1$ and $2$. As in other photon storage scenarios \cite{S2}, we will replace the field operators in the above dynamical equations by their mean values to see how their profiles evolve under given conditions. The dynamical equations for the field profiles in Eqs. (1)-(3) of the main text are thus obtained by averaging out the quantum noise operators, $\langle \hat{\zeta}_l({\bf x},t)\rangle=0$.

\subsection*{B.~~~~~~More examples of pulse evolution under realistic interaction}
\renewcommand{\theequation}{B-\arabic{equation}}
\setcounter{equation}{0}

\begin{figure}[t!]
\vspace{-0cm}
\centering
\epsfig{file=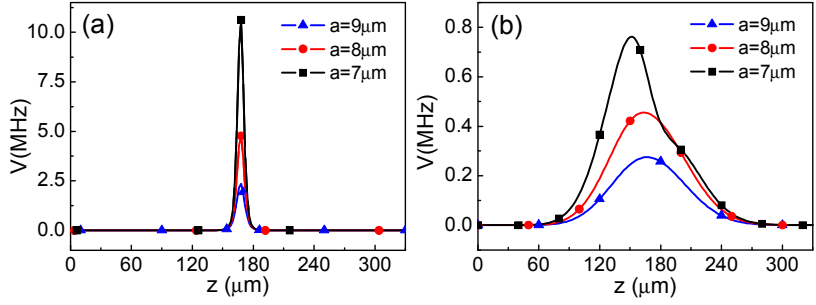,width=0.6\linewidth,clip=}
{\vspace{-0.2cm}
\caption{ Comparison between the potential from a point particle (a) and that due to a stopped distributing spinwave (b) for different ensemble separations $a$. Here the potentials originate from a vdW potential $\Delta({\bf x}-{\bf x}')$ in Eq. (\ref{interaction}).  The profiles in (b) from stopping two counter-propagating pulses are achieved with a control field of $\Omega_{c}^{M}=2.0\times2\pi$ MHz, $\tau_{c}=8\mu$s, and $t_{c}=40$ $\mu$s. When these obtained spinwaves shrink to one point on the ensemble axis, they give the potentials in (a). The photons resonantly coupled to the energy levels have $\Omega_{p}^{M}=0.01$MHz, $t_{p}=12.0$ $\mu$s and duration $\tau_{p}=7.0$ $\mu$s. The atomic level scheme is the same as the one in Figs. 2-4 of the main text, and the atomic density of the $0.335$-mm long ensembles with their diameters $d=2$ $\mu$m is $N=2\times10^{13}$ cm$^{-3}$.   }}
\vspace{-0cm}
\end{figure}

\begin{figure}[b!]
\vspace{-0cm}
\centering
\epsfig{file=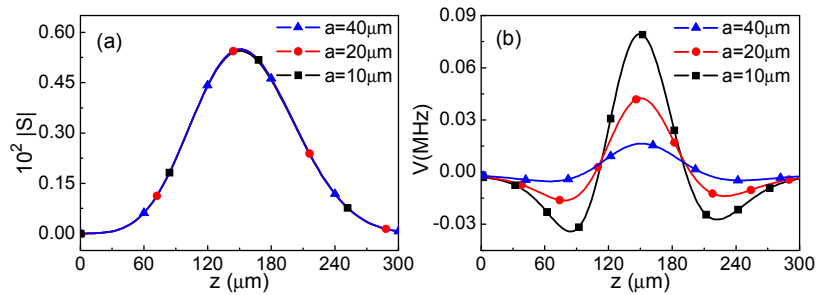,width=0.6\linewidth,clip=}
{\vspace{-0.2cm}
\caption{ Realized dipole-dipole potentials from $\Delta({\bf x}-{\bf x}')\sim C_3/|{\bf x}-{\bf x}'|^3$ in Eq. (\ref{interaction}). These potentials are created by applying a static electric field parallel to the ensemble axis, after the photons are stopped and converted to Rydberg spinwave packets. The atomic level scheme is the same as the one in Figs. 2-4 of the main text, to have $\left\vert C_{3}\right\vert \approx 6.65\times10^{5}$ MHz $\mu$m$^{3}$. The other parameters of the $0.3$-mm long ensembles are the same as those in Fig. S-1. Due to the large ensemble separations, the accompanying vdW potentials almost take no effect, 
resulting in the same spinwave profiles as seen in (a).}}
\vspace{-0cm}
\end{figure}

The interaction potential a pulse experiences in the concerned processes is determined by the spatial extension of the other Rydberg spinwave packet, as indicated by Eq. (\ref{inter}) or Eq. (4) of the main text, which comes from the full dynamics discussed in the above. This leads to a significant difference from the interaction with a point Rydberg atom. We illustrate the fact with the comparison in Fig. S-1. The finally obtained realistic potential profiles in the illustrated situations correspond to those of the standstill Rydberg spinwaves. Though the extending spinwave packets lower the potential magnitudes, the potential values distribute less abruptly over the ensembles. Such potentials due to wave packets instead of point particles exist for any type of interaction. For instance, the dipole-dipole interaction potentials between the stored wave packets should also be evaluated in this way, and are found to be more inhomogeneous than those of vdW interaction; see Fig. S-2. The dipole-dipole interaction is applied in another photonic gate proposal based on stored photon pulses \cite{S3}. More realistic considerations should be therefore included in quantum information processing by the similar methods. 

One of the meaningful consequences of such realistic interaction potentials is that they lead to different dissipation rates over the pulses. In Ref. \cite{S4} about another topic related to Rydberg EIT, the interaction between pulses is also modeled as the one between extending wave packets, but their dissipation is assumed to be with a global rate independent of the pulse distributions. The exact details of pulse dissipation should be clarified by approaching the processes with the complete dynamics of the quantum field profiles. Here we present two more examples to show how such space-time dependent dissipation manifest in photon storage. The first one
in Fig. S-3 is about the even closer ensembles in Fig. 3(a1) of the main text. The interaction between the pulses in these setups becomes stronger. Since the counter-propagating pulses have spatial extensions, the interaction between their fronts is intense before they are stopped in the medium. An extreme result is that half of a pulse will be absorbed. Negligible decay of the Rydberg levels after the control field is turned off maintains stable asymmetric spinwave profiles during the storage period. The other example in Fig. S-4 illustrates the dynamical evolutions of two co-propagating photon pulses and their associated spinwaves before they are stopped together. Here we use a narrower pulse bandwidth (corresponding to longer pulse durations $\tau_p$) so that the pulse dissipation is almost completely due to their interaction. This example clearly demonstrates the significant effects of pulse interaction. Due to a longer interaction time, only a small portion of the initially induced spinwaves will remain in the end.

\begin{figure}[h!]
\vspace{1cm}
\centering
\epsfig{file=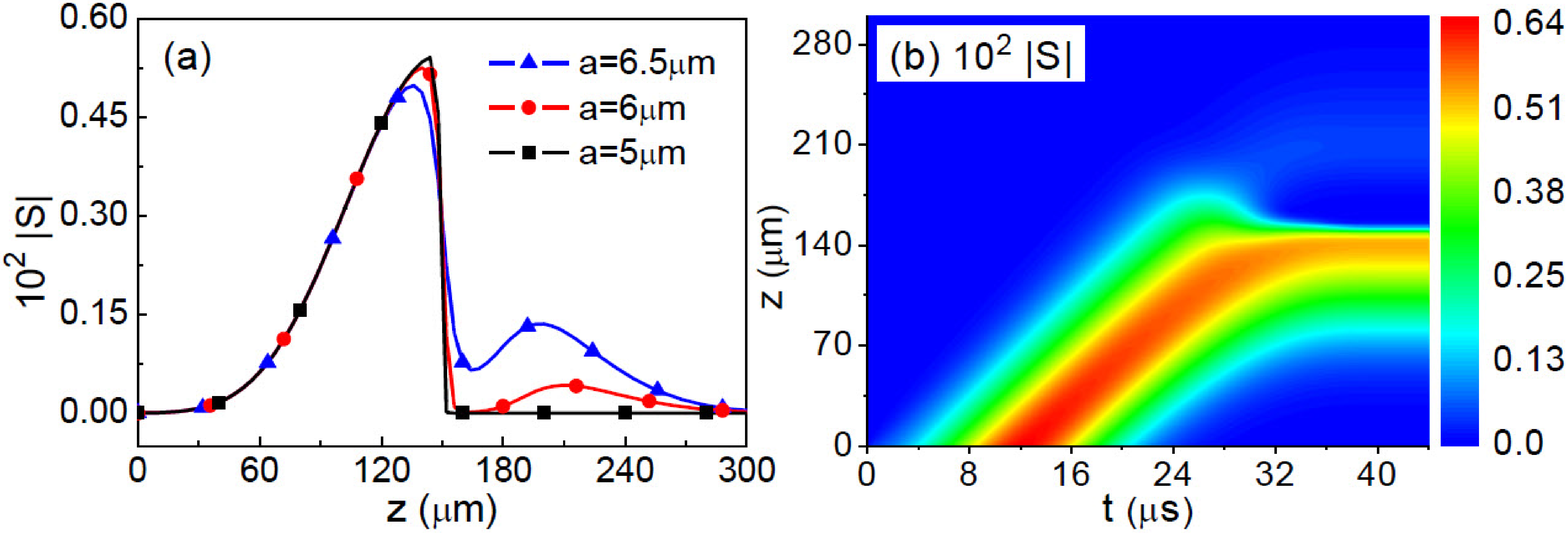,width=0.6\linewidth,clip=}
{\vspace{-0.2cm}
\caption{ (a) Created spinwave profiles for various relatively small ensemble separations. (b) Dynamical evolution of the associated spinwave field on the axis of ensembles. Here the ensemble separation is $a=6$ $\mu$m. The control field is switch off at a speed of
$\tau_{c}=10$ $\mu$s, and ensemble length is $0.3$ mm. The other parameters are the same as Fig. S-1.  }}
\vspace{-0cm}
\end{figure}
\begin{figure}[h!]
\vspace{-0cm}
\centering
\epsfig{file=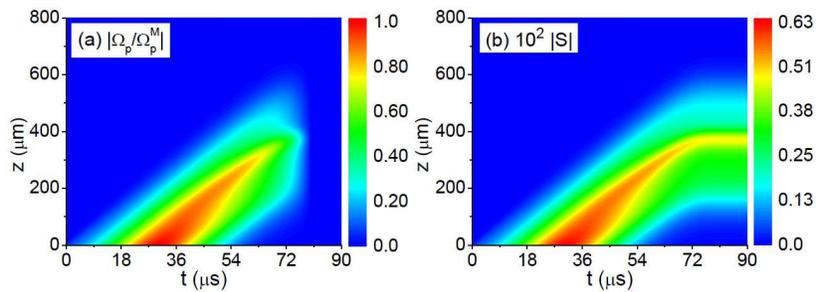,width=0.6\linewidth,clip=}
{\vspace{-0.2cm}
\caption{ (a) Dynamics of two co-propagating photon pulses during their storage. (b) Dynamics of the associated spinwaves.
Here we take $a=10$ $\mu$m, $t_{p}=30.0$ $\mu$s, $\tau_{p}=18.0$ $\mu$s, $\tau_{c}=10\mu s$, and $t_{c}=80$ $\mu$s. The other parameters
are the same as those as in Fig. S-1. }}
\vspace{-0cm}
\end{figure}

\subsection*{C.~~~~~~Further discussion on non-adiabatic corrections}
 \renewcommand{\theequation}{C-\arabic{equation}}
 \setcounter{equation}{0}

 \begin{figure}[h!]
\vspace{-0cm}
\centering
\epsfig{file=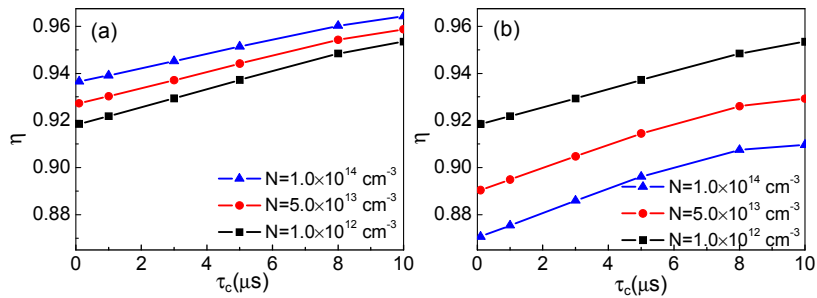,width=0.6\linewidth,clip=}
{\vspace{-0.2cm}
\caption{ (a) Relations between the storage efficiency $\eta=|S^{M}(t_s)/S^{M}(t_{in})|$ and the control field switch speed indicated
by $\tau_{c}$, where $|S^{M}(t_s)|$ and $|S^{M}(t_{in})|$ are, respectively, the peak value of the stored and initial spinwave. These results for different atomic densities $N$ are about the storage processes without pulse interaction. (b) Corresponding relations 
for two co-propagating pulses in the ensembles separated by a distance $a=13$ $\mu$m. In both situations we take $\tau_{p}=10$ $\mu$s, $t_{p}=20$ $\mu$s, and $t_{c}=40$ $\mu$s. The ensemble length is $6$ mm. The other
parameters are the same as those in Fig. S-1. }}
\vspace{-0cm}
\end{figure}

We now take a further look at the effects from a fast-changing Rabi frequency $\Omega_c(t)$ of control field. To neglect the photon losses from these effects, the systems should evolve along adiabatic passages during storage and retrieval of photons. The condition for realizing such adiabatic passages can be formulated in terms of a change rate of the mixing angle $\vartheta (t)=\arctan g\sqrt{N}/\Omega_c(t)$ as $\dot{\vartheta}\ll \frac{1}{\hbar}|E_{\pm}-E_0|$, where $|E_{\pm}-E_0|$ is the energy difference between bright-state and dark-state polaritons (see, e.g. \cite{S5}). The change rate, which takes the form
\begin{eqnarray}
\frac{d}{dt}\vartheta(t)=-\frac{g\sqrt{N}}{g^2N+\Omega_c^2(t)}\dot{\Omega}_c(t),
\end{eqnarray}
scales inversely with the square root of the atomic density $N$ in our concerned regime realizing slow light
($g\sqrt{N}/|\Omega_c(t)|\gg 1$). In ensembles of high density $N$, insignificant non-adiabatic effects are therefore expected.
This is seen from Fig. S-5(a) about the processes without the interaction between pulses, in which the storage efficiency goes up with the density and the difference between fast and slow switching of a control field is not so large. Meanwhile, in the presence of the pulse interaction as shown in Fig. S-5(b), the tendency of storage efficiency with atomic density will be totally reversed. The interaction between spinwaves becomes stronger in an ensemble of higher density, because the spinwave profiles $S({\bf x},t)=\sqrt{N}\langle\hat{\sigma}_{gr}\rangle ({\bf x},t)$ are magnified by the square root of $N$. Accordingly more significant dissipation will exist in the ensemble. These results demonstrate that the dissipation from non-adiabatic corrections is hardly comparable with that due to pulse interaction. Stopping and regenerating photons in the setups can be done with quickly switching control field too.

\end{widetext}


\begin{thebibliography}
\vspace{}
\bibitem {EIT1} J. D. Pritchard, D. Maxwell, A. Gauguet, K. J. Weatherill, M. P. A. Jones, and C. S. Adams, Phys. Rev. Lett. 105, 193603 (2010).
\bibitem {EIT0} H. Schempp, G. G\"{u}ter, C. S. Hofmann, C. Giese, S. D. Saliba, B. D. DePaola, T. Amthor, M. Weidem\"{u}ller, S. Sevin\c{c}li, and T. Pohl, Phys. Rev. Lett. 104, 173602 (2010).
\bibitem {rv-eit} M. Fleischhauer, A. Imamoglu, and J. P. Marangos, Rev. Mod. Phys. 77, 633 (2005).
\bibitem {EIT-0} C. Ates, S. Sevin\c{c}li, and T. Pohl, Phys. Rev. A 83, 041802(R) (2011).
\bibitem {EIT-1} S. Sevin\c{c}li, N. Henkel, C. Ates, and T. Pohl, Phys. Rev.
Lett. 107, 153001 (2011).
\bibitem {EIT-2} D. Petrosyan, J. Otterbach, and M. Fleischhauer, Phys. Rev. Lett. 107, 213601 (2011).
\bibitem {EIT-3} J. D. Pritchard, C. S. Adams, and K. M{\o}lmer, Phys. Rev. Lett. 108, 043601 (2012).
\bibitem {EIT-4} M. G\"{a}rttner and J. Evers, Phys. Rev. A 88, 033417 (2013).
\bibitem {EIT-5}J. Stanojevic, V. Parigi, E. Bimbard, A. Ourjoumtsev, and P. Grangier, Phys. Rev. A 88, 053845 (2013).
\bibitem {EIT-6} Y.-M. Liu, D. Yan, X.-D. Tian, C.-L. Cui, and J.-H. Wu, Phys. Rev. A 89, 033839 (2014).
\bibitem {EIT-7} W. Li, D. Viscor, S. Hofferberth, and I. Lesanovsky, Phys. Rev. Lett. 112, 243601 (2014).
\bibitem {EIT-8} H. Wu, M.-M. Bian, L.-T. Shen, R.-X. Chen, Z.-B. Yang, and S.-B. Zheng, Phys. Rev. A 90, 045801 (2014).
\bibitem {EIT-a} M. G\"{a}rttner, S. Whitlock, D. W. Sch\"{o}nleber, and J. Evers, Phys. Rev. Lett. 113, 233002 (2014).
\bibitem {EIT-9} D. Viscor, W. Li, and I. Lesanovsky, New J. Phys. 17, 033007 (2015).
\bibitem {EIT-10} Y.-M. Liu, X.-D. Tian, D. Yan, Y. Zhang, C.-L. Cui, and J.-H. Wu, Phys. Rev. A 91, 043802 (2015).
\bibitem {bd} P. Bienias, S. Choi, O. Firstenberg, M. F. Maghrebi, M. Gullans, M. D. Lukin, A. V. Gorshkov, and H. P.
B\"{u}chler, Phys. Rev. A 90, 053804 (2014).
\bibitem {lt} J. Otterbach, M. Moos, D. Muth, and M. Fleischhauer, Phys. Rev. Lett. 111, 113001 (2013).
\bibitem {Ex1} T. Peyronel, O. Firstenberg, Q.-Y. Liang, S. Hofferberth, A. V. Gorshkov, T. Pohl, M. D. Lukin, and V. Vuleti\'{c},
Nature (London) 488, 57 (2012).
\bibitem {Ex2} Y. O. Dudin, F. Bariani, and A. Kuzmich, Phys. Rev. Lett. 109 (2012).
\bibitem {Ex3} C. S. Hofmann, G. G\"{u}nter, H. Schempp, M. Robert-de-Saint-Vincent, M. G\"{a}rttner, J. Evers, S. Whitlock, and
M. Weidem\"{u}ller, Phys. Rev. Lett. 110, 203601 (2013).
\bibitem {Ex4} O. Firstenberg, T. Peyronel, Q.-Y. Liang, A. V. Gorshkov, M. D. Lukin, and V. Vuleti\'{c},
Nature (London) 502, 71 (2013).
\bibitem {Ex5} D. Maxwell, D. J. Szwer, D. Paredes-Barato, H. Busche, J. D. Pritchard, A. Gauguet, M. P. A. Jones, and C. S. Adams,
Phys. Rev. A 89, 043827 (2014).
\bibitem {g1} I. Friedler, D. Petrosyan, M. Fleischhauer, and G. Kurizki, Phys. Rev. A 72, 043803 (2005).
\bibitem {g2} B. He, A. MacRae, Y. Han, A. Lvovsky, and C. Simon, Phys. Rev. A 83, 022312 (2011).
\bibitem {g3} E. Shahmoon, G. Kurizki, M. Fleischhauer, and D. Petrosyan, Phys. Rev. A 83, 033806 (2011).
\bibitem {g4} A. V. Gorshkov, J. Otterbach, M. Fleischhauer, T. Pohl, and M. D. Lukin, Phys. Rev. Lett. 107, 133602 (2011).
\bibitem {g5} B. He, A. V. Sharypov, J. Sheng, C. Simon, and M. Xiao, \prl 112, 133606 (2014).
\bibitem {c1} A. Rispe, B. He, and C. Simon, \prl 107, 043601 (2011).
\bibitem {c2} C. Vo, S. Riedl, S. Baur, G. Rempe, and S. D\"{u}rr, Phys. Rev. Lett. 109, 263602 (2012).
\bibitem {p1} D. Paredes-Barato and C. S. Adams, Phys. Rev. Lett. 112, 040501 (2014).
\bibitem {p2} M. Khazali, K. Heshami, and C. Simon, Phys. Rev. A 91, 030301(R) (2015).
\bibitem {sw} S. Baur, D. Tiarks, G. Rempe, and S. D\"{u}rr, Phys. Rev. Lett. 112, 073901 (2014).
\bibitem {tr1} D. Tiarks, S. Baur, K. Schneider, S. D\"{u}rr, and G. Rempe, Phys. Rev. Lett. 113, 053602 (2014).
\bibitem {tr2} H. Gorniaczyk, C. Tresp, J. Schmidt, H. Fedder, and S. Hofferberth, Phys. Rev. Lett. 113, 053601 (2014).
\bibitem {d1} Y. O. Dudin and A. Kuzmich, Science 336, 887 (2012).
\bibitem {d2} D. Maxwell, D. J. Szwer, D. Paredes-Barato, H. Busche, J. D. Pritchard, A. Gauguet, K. J. Weatherill, M. P. A. Jones, and C. S. Adams, Phys. Rev. Lett. 110, 103001 (2013).
\bibitem {f-l}M. Fleischhauer and M. D. Lukin, \pra 65, 022314 (2001).
\bibitem {ad0} D. M{\o}ller, L. B. Madsen, and K. M{\o}lmer, Phys. Rev. Lett. 100, 170504 (2008).
\bibitem {ad1} D. Petrosyan and K. M{\o}lmer, Phys. Rev. A 87, 033416 (2013).
\bibitem {ad2} J. Qian, J. Zhai, L. Zhang, and W. Zhang, Phys. Rev. A 91, 013411 (2015).
\end{thebibliography}

\begin{thebibliography}{99}
\renewcommand{\thebibliography}{S-\arabic{bibliography}}
\vspace{-0cm}
\bibitem [S1] {S1} C. W. Gardiner and P. Zoller, {\it Quantum Noise} (Springer-Verlag, Berlin, 2000).
\bibitem [S2] {S2} A. V. Gorshkov, A. Andre, M. D. Lukin, and A. S. S{\o}rensen, Phys. Rev. A 76, 033805 (2007).
\bibitem [S3] {S3} D. Paredes-Barato and C. S. Adams, Phys. Rev. Lett. 112, 040501 (2014).
\bibitem [S4] {S4} B. He, A. V. Sharypov, J. Sheng, C. Simon, and M. Xiao, \prl 112, 133606 (2014).
\bibitem [S5] {S5} D. Petrosyan and K. M{\o}lmer, Phys. Rev. A 87, 033416 (2013).
\end{thebibliography}
\end{document}